\documentclass[final,3p,times,twocolumn]{elsarticle}
 \biboptions{comma,sort&compress}
\usepackage{here}
 \usepackage{graphicx}
  \usepackage{epsfig}

\def\nin{\noindent}
\def\beq{\begin{equation}}
\def\eeq{\end{equation}}
\def\bea{\begin{eqnarray}}
\def\eea{\end{eqnarray}}

\def\mZ{\ensuremath{M_Z}}
\def\as{\ensuremath{\alpha_{\scriptscriptstyle S}}}
\def\asZ{\ensuremath{\as(\mZ^2)}}

\def\pt{\ensuremath{p_{\rm T}}}

\def\absy{\ensuremath{|y|}}

\def\AKT{anti-${k_t}$}
\def\nlojet{NLOJET++} 
\def\applgrid{APPLGRID}

\def\asNomVec{\ensuremath{\bar{\alpha}_{\scriptscriptstyle S}^{\rm Nom}}}

\def\asAvVec{\ensuremath{\bar{\alpha}_{\scriptscriptstyle S}^{\rm Av}}}
\def\asAv{\ensuremath{\alpha_{\scriptscriptstyle S}^{\rm Av}}}

\def\chiSq{\ensuremath{\chi^2}}

\journal{Nuc. Phys. (Proc. Suppl.)}

\begin{document}

\begin{frontmatter}



\title{Evaluation of \as~using the ATLAS inclusive jet cross-section data}

\author{B.~Malaescu}
\address{CERN, CH--1211, Geneva 23, Switzerland}

\begin{abstract}
\noindent

We present a determination of the strong coupling constant using ATLAS inclusive jet cross section data at $\sqrt{s}=7{\rm TeV}$,
with their full information on the bin-to-bin correlations.
Several procedures for combining the statistical information from the data inputs are compared.
The theoretical prediction is obtained using NLO QCD with non-perturbative corrections.
Our determination uses inputs with transverse momenta between 45 and 600~GeV, the running of the strong coupling being also tested in this range.
Good agreement is observed when comparing our result with the world average at the Z-boson scale, as well as with the most recent results from the Tevatron.

\end{abstract}

\begin{keyword}

QCD \sep strong coupling \sep jets


\end{keyword}

\end{frontmatter}


\section{Introduction}
\nin
In the Standard Model of particle physics~(SM), the strong coupling constant \as~is one of the fundamental parameters.
Testing the energy dependence~(running) of \as~over a wide range provides an implicit test of QCD and probes potential effects from ``New Physics''.

We present a determination of \as~using the ATLAS inclusive jet cross section data~\cite{Aad:2011fc}, allowing for the first test of the running of \as~up to the TeV scale.
For a given \as~value, the perturbative NLO QCD prediction of the inclusive jet cross section is computed, and is then corrected for non-perturbative effects.
The perturbative computation uses the NLO QCD proton parton density functions~(PDFs), as obtained by various groups from other independent data sets.
We first determine \asZ~using the measured cross section in each data bin, and then propagate the experimental uncertainties and bin-to-bin correlations.
Finally, an unbiased averaging procedure is used and the uncertainties are propagated to the corresponding result.

The results presented here are based on the studies described into detail in \cite{Malaescu:2012ts}.

\section{Input Data}
\label{Sec:InputData}
\nin

In this study we exploit the unfolded double-differential ATLAS inclusive jet cross section data, in bins of the absolute rapidity 
$(\absy)$ and as a function of the jet transverse momentum~(\pt), with the statistical and systematic uncertainties, and their correlations~\cite{Aad:2011fc}.

This measurement, performed for \AKT~jets~\cite{Cacciari:2008gp} with a distance parameter R=0.4 and R=0.6 respectively, covers a large phase-space region going
up to 4.4 in rapidity and from 20~GeV to more than 1~TeV in \pt.
The detailed analysis of the uncertainties and their correlations is one of the important achievements of this measurement.

The dominant uncertainty in this measurement is due to the jet energy scale~(JES) calibration.
Other smaller systematic uncertainties are due to the luminosity measurement, multiple proton-proton interactions, trigger efficiency, jet reconstruction and 
identification, jet energy resolution and deconvolution of detector effects.
The amplitudes of these (asymmetric)~uncertainties and their (anti-)~correlations between the various \absy~and \pt~bins 
are provided using a set of 87 independent nuisance parameters.
Depending on the rapidity bin, the systematic uncertainties account for $20-60\%$ at low \pt, and $20-40\%$ at high \pt, while they are smaller for
intermediate jet transverse momentum values.

The statistical uncertainties and their (anti-)~correlations are provided through a covariance matrix for each rapidity bin.
They are in general much smaller than the systematics, except for the high \pt~region, where they can be larger.

It has been pointed out in~\cite{Aad:2011fc} that, when performing the comparison between data and the theoretical prediction, somewhat larger differences are observed for R=0.4 than for R=0.6 jets.
In our study, we do not perform a simultaneous treatment of the two jet sizes, because the correlations of the uncertainties of these measurements~(expected to be large) are not available.
We use the measurement of jets with R=0.6 for determining our nominal result, while the comparison with the result using jets with R=0.4 provides a systematic uncertainty.

\section{Theoretical Prediction}
\label{Sec:Theory}
\nin

Theoretical predictions for the inclusive jet cross sections are calculated in perturbative~QCD at next-to-leading order~(NLO)  using the \nlojet~\cite{Nagy:2003tz} program.
The effects of hadronisation and underlying event activity on jet production have been studied using Monte Carlo generators. 
Each bin of the perturbative cross section is multiplied by the non-perturbative corrections to obtain the prediction at the particle level that can be compared to the data.

To allow for fast and flexible evaluation of the cross section using different values of $\as$~(exploiting the \as~scans performed in the PDF fits),
as well as for evaluation of the PDF and scale uncertainties, the \applgrid~\cite{Carli:2010rw} software was interfaced with \nlojet~in order to calculate the perturbative coefficients once
and store them in a look-up table.
The nominal theoretical prediction is derived with the CT10~\cite{Lai:2010vv} NLO parton density functions, the corresponding uncertainties~(eigenvectors) being propagated too.
Uncertainties due to the choise of the PDF set are evaluated by comparing the nominal results with the ones obtained with MSTW~2008~\cite{Martin:2009iq},
NNPDF~2.1~\cite{Ball:2011mu, Ball:2011uy} and  HERAPDF~1.5~\cite{HERAPDF15}.

In our analysis we have used the non-perturbative corrections and uncertainties evaluated in~\cite{Aad:2011fc}, using leading-logarithmic parton shower generators with their various tunes.
For $R=0.4$ the correction factors are dominated by the effect of hadronisation and are approximately 0.95 at jet $\pt=20$~GeV, increasing closer to unity at higher $\pt$.
For $R=0.6$ the correction factors are dominated by the UE and are approximately 1.6 at jet $\pt=20$~GeV decreasing to between $1.0-1.1$ for jets above $\pt=100$~GeV.

\section{\as~Evaluation Procedure}
\label{Sec:EvaluationProcedure}
\nin

The determination of \asZ~goes through exploiting the experimental information in individual (\pt; \absy) bins of the jet cross section, together with the propagation of
the (asymmetric)~uncertainties and bin-to-bin correlations.

An \as~value is obtained in each bin of the measurement, using the theoretical prediction and the experimental cross section.
The theoretical prediction provides, in each (\pt; \absy) bin, a one-to-one mapping between \as~and the inclusive jet cross section.
The nominal value of the measured cross section is mapped by this procedure to the nominal \as~value in the corresponding bin.

A series of pseudo-experiments, each of them including a complete set of cross section values in all the (\pt; \absy) bins, are generated using the
experimental input described in section~\ref{Sec:InputData}.
The statistical fluctuations are generated using the information provided by the covariance matrices.
Each nuisance parameter for a given component of the systematic uncertainties is treated as $100\%$ correlated between all the bins.
The various nuisance parameters are treated as independent with respect to each other.
The distributions of the experimental uncertainties are modelled using asymmetric Gaussian shapes, centered at zero, with $50\%$ probability of generating 
pseudo-measurements at negative or positive values.
Other possible models of these distributions have been tested, yielding similar results.
For each pseudo-experiment, the same procedure as for the nominal measurements is applied to obtain an \as~value in each bin.
In view of the study of various possible averaging procedures, it is interesting to determine~(after a small symmetrisation of uncertainties) the
covariance matrix of the \as~values.

In order to avoid potential biases introduced by some averaging procedures, several options for performing the combination of the inputs in the various (\pt; \absy) bins have been tested.
All of them aim at obtaining an average value of the strong coupling constant~(\asAv).

The first procedure consists of a simple average, where all the input \as~values have the same weight.
The $2^{nd}$ procedure which has been tested consists of a weighted average, where the weights of the input \as~values are proportional to the inverse of their squared total uncertainties.
Although the final uncertainties on these two averages might not be optimal, the weights of the input \as~values are well behaved~(i.e.
they are all in the $[0;1]$ interval, with the sum equal to unity).
The $2^{nd}$ method also has the advantage that more precise contributions get larger weights.

The $3^{rd}$ averaging procedure consists of the minimisation of a ``standard'' \chiSq~with correlations
\bea
   \chiSq = (\asNomVec - \asAvVec) \cdot C^{-1} \cdot (\asNomVec - \asAvVec)^{T},
   \label{Eq:chi2TotCov}
\eea
with respect to \asAv.
Here, \asNomVec~is the vector of nominal \as~values~(obtained in individual (\pt; \absy) bins) entering the average, $C$ is their covariance matrix,
while \asAvVec~is a vector containing values equal to \asAv~(to be fitted) and having the same size as \asNomVec.

It has been shown in~\cite{D'Agostini:1993uj} that this approach, although very commonly used, can yield biased average values.
Actually, in presence of (not very well understood)~strong correlations among the inputs, the average value can be even outside the range of the input values, which seems unacceptable.
Indeed, the minimisation of the function in Eq.~\ref{Eq:chi2TotCov}, with respect to \asAv~(equivalent to the minimisation of the variance of the output of a weighted average, cf. Gauss-Markov
theorem), yields
\bea
   \asAv = \frac{\bar{1} \cdot C^{-1} \cdot (\asNomVec)^{T}}{\bar{1} \cdot C^{-1} \cdot \bar{1}^{T}},
   \label{Eq:chi2minimum}
\eea
where $\bar{1}$ is a vector with all the entries equal to unity and the same size as \asNomVec.
Just as in the two previous methods, the average resulting from the \chiSq~minimisation is a weighted mean of the input values.
However, in presence of strong correlations, these weights can even be negative or larger than unity.

The explanation of the bias in this procedure, as provided in~\cite{D'Agostini:1993uj}, makes responsible the input covariance matrix because of ``the linearisation on which the usual error propagation relies''.
In~\cite{Blobel:2003wa}, the bias was explained by the use of the ill-behaved inversion of the covariance matrix,
as well as the treatment of all the uncertainties as absolute~(i.e. corresponding to additive effects),
whereas at least some of them should be treated as relative uncertainties~(i.e. corresponding to multiplicative effects).
In \cite{Malaescu:2012ts} we have pointed out one additional problem:
this combination procedure fully relies on the knowledge of the bin-to-bin correlations~(difficult to check in a particular measurement),
in order to derive ``optimal weights'', providing the smallest uncertainty on the average.
This yields too optimistic results, as the uncertainties on the shape and correlations for each component of the systematic uncertainties should also be taken into account.

The three combination procedures described above, make use of weighted averages.
They distinguish between each other by the way how the weights of the inputs in the average are obtained.
Another possibility that we have considered for performing the combination consists in a geometrical average.

\section{Experimental results}
\label{Sec:Results}
\nin

As explained in the previous section, a nominal \as~value has been determined in each (\pt; \absy) bin, while the statistical and systematic
uncertainties, together with their correlations, have been propagated using pseudo-experiments.
For very low and very large \pt~values, the total data uncertainties are relatively large, and a fluctuation of the cross section by $\pm 1\sigma$~(or more)
yields \as~values for which the theoretical prediction is not reliable.
It is for this reason that we discard the \pt~bins with large experimental uncertainties~(hence small weights) in the final \as~average computation,
and we estimate a systematic uncertainty due to this choice.
A number of 42 (\pt; \absy) bins are used in the computation of the various averages.

The precision of the weighted average~(WA) computed in the individual \absy~intervals of the experimental measurement~\cite{Aad:2011fc},
is similar to the precision of the values obtained in the corresponding~(input) \pt~bins~\cite{Malaescu:2012ts}.
This is due to the strong correlations between~(and the similar size of) the uncertainties in the \pt~bins used for the average inside a given \absy~bin.
Performing the combination of all the 42 (\pt; \absy) bins, one obtains
\bea
   \as^{\rm WA} = 0.1151^{+0.0047}_{-0.0047}.
\label{Eq:NominalWeightedAverage}
\eea
The experimental uncertainty of this average is about $30\%$ smaller compared to the one obtained in the most precise individual \absy~bin~(i.e. the central bin).
This gain in precision is due to the rather small correlations of the systematic uncertainties, between the central and the forward region.
The results~(central values and uncertainties) obtained using the weighted average, the simple average
and the geometrical mean are consistent between each other, as expected for a set of inputs with similar uncertainties.

The result of the weighted average, together with the covariance matrix of the uncertainties for the \as~values,
have been inserted into Eq.~\ref{Eq:chi2TotCov}, yielding a $\chi^{2}/{\rm dof}=0.54$~(for $41$ degrees of freedom).
This represents an implicit test of the RGE, used for the perturbative prediction as well as the fits for determining the PDFs used here.
Very reasonable values for the $\chi^{2}/{\rm dof}$ are obtained in the individual \absy~bins too.
Taking into account the bin-to-bin correlations is very important for the determination of these $\chi^2$ values.
Ignoring the correlations would actually yield $\chi^2$ values smaller by more than one order of magnitude.

The procedure based on the minimisation of a $\chi^2$ with correlation~(see Eq.~\ref{Eq:chi2TotCov}) yields
\bea
   \as^{ \chi^2_{\rm min}} = 0.1165^{+0.0033}_{-0.0033}.
\eea
Compared to the previous averaging procedures, this result has a smaller experimental uncertainty and slightly smaller $\chi^{2}/{\rm dof}$~($0.53$), as expected.
However, the central value obtained here is questionable, since the weights of this average strongly rely on the exact knowledge of the bin-to-bin correlations and are not well behaved.
Indeed, a large fraction~(almost half) of the weights of the individual \pt~bins in this average are smaller than zero.
Looking at the similar averages computed in individual \absy~bins, they are in the range between $0.1100$ and $0.1243$, several of them being outside the range of the corresponding input values.
This is due to the same bad behaviour of the weights, some of which are negative, while some others are larger than one.

It is due to the issues observed for the minimisation of a $\chi^2$ with correlation, compared to the well behaved weighted average,
that we choose to use the value obtained with the weights given by the inverse of the squared total uncertainties, for our nominal result.

\begin{figure}[hbt] 
\centerline{\includegraphics[width=7.5cm]{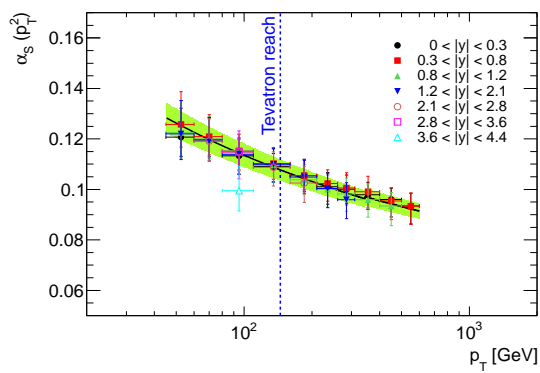}}
\caption{
  Weighted $\as$ average~(green band) evolved to the corresponding $p_{\rm T}$ scale, together with the values obtained in all the (\pt; \absy) bins
  used in the combination.
  The blue vertical line indicates the highest \pt~value used in the Tevatron \asZ~determination, from the inclusive jet cross section.
}
\label{Fig:asRunning}
\end{figure} 
\nin

The nominal result for \as~evolved by ${\rm N}^{3}{\rm LO}$ RGE running, its uncertainty, and the comparison with the inputs used in the combination are shown in Fig.~\ref{Fig:asRunning}.
We also indicate the highest \pt~value~($145$~GeV) used in the \asZ~determination, from the inclusive jet cross section, at Tevatron~\cite{Abazov:2009nc}.
Our combination procedure uses a \pt~range going up to four times higher.
We also use inputs from the forward region, up to $\absy = 4.4$, while the Tevatron determination is restricted to $\absy < 1.6$.

Results with a similar precision are obtained using \AKT~jets with R=0.4.
The central value of the corresponding weighted average is, however, shifted downwards with respect to our nominal result~(from \AKT~jets with R=0.6), by about $0.0060$.
This is expected, due to a systematic difference seen between the comparisons of the two (strongly correlated)~measurements with the corresponding theoretical predictions~(see
section~\ref{Sec:InputData}).
The difference between the \as~determinations with the two jet sizes is actually our largest uncertainty and an improvement in the understanding of these differences is desirable.

\section{Conclusions}
\label{Sec:Conclusions}
\nin

We have performed a determination of the strong coupling constant at the Z scale, using the ATLAS inclusive jet cross section data.
Our final result accounts for~\cite{Malaescu:2012ts} :
\bea
   &&\!\!\!\!\!\!\!\!\!\!\!\!\!\!\!\!\!\!\!\!\!\!\!\!
      \asZ = 0.1151  \pm 0.0001~({\rm stat.})  \pm 0.0047~({\rm exp.~syst.}) \nonumber \\
   && \pm 0.0014~({\rm\pt~range})  \pm 0.0060~({\rm jet~size}) \nonumber \\
   && ^{+ 0.0044}_{-0.0011}~({\rm scale}) ^{+0.0022}_{-0.0015}~({\rm PDF~choice}) \nonumber \\
   && \pm 0.0010~({\rm PDF~eig.})  ^{+0.0009}_{-0.0034}~({\rm NP~corrections}) , \nonumber
\eea
where all the statistical and systematic uncertainties are included.
Our value is in good agreement with the latest (preliminary) update of the \asZ~world average~($0.1183 \pm 0.0010$)~\cite{Bethke:2011tr}, as well as with the latest result from a hadron-hadron
collider~($0.1161^{+0.0041}_{-0.0048}$)~\cite{Abazov:2009nc}.
Although our result is less precise, it includes for the first time the measurements of the inclusive jet cross section up to $600$~GeV.
The running of \as~has also been tested, in the \pt~range between $45$ and $600$~GeV, and no evidence of a deviation from the QCD prediction has been observed.

\section*{Acknowledgements}
\nin
The author thanks the organizers for the invitation and acknowledges his fruithful collaboration with P.~Starovoitov.





\begin{thebibliography}{999}
\vspace*{-0.25cm}


\bibitem{Aad:2011fc}
  G.~Aad {\it et al.}  [ATLAS Collaboration],
  arXiv:1112.6297 [hep-ex].

\bibitem{Malaescu:2012ts} 
  B.~Malaescu and P.~Starovoitov,
  Eur.\ Phys.\ J.\ C {\bf 72}, 2041 (2012)
  [arXiv:1203.5416 [hep-ph]].

\bibitem{Cacciari:2008gp}
  M.~Cacciari, G.~P.~Salam and G.~Soyez,
  JHEP {\bf 0804} (2008) 063
  [arXiv:0802.1189 [hep-ph]].

\bibitem{Nagy:2003tz}
  Z.~Nagy,
  Phys.\ Rev.\ D {\bf 68} (2003) 094002
  [hep-ph/0307268].

\bibitem{Carli:2010rw}
  T.~Carli {\it et al.},
  Eur.\ Phys.\ J.\ C {\bf 66} (2010) 503
  [arXiv:0911.2985 [hep-ph]].

\bibitem{Lai:2010vv}
  H.~-L.~Lai {\it et al.},
  Phys.\ Rev.\ D {\bf 82} (2010) 074024
  [arXiv:1007.2241 [hep-ph]].

\bibitem{Martin:2009iq}
  A.~D.~Martin, W.~J.~Stirling, R.~S.~Thorne and G.~Watt,
  Eur.\ Phys.\ J.\  C {\bf 63} (2009) 189
  [arXiv:0901.0002 [hep-ph]].


\bibitem{Ball:2011mu}
  R.~D.~Ball, V.~Bertone, F.~Cerutti, L.~Del Debbio, S.~Forte, A.~Guffanti, J.~I.~Latorre and J.~Rojo {\it et al.},
  Nucl.\ Phys.\ B {\bf 849} (2011) 296
  [arXiv:1101.1300 [hep-ph]].

\bibitem{Ball:2011uy}
  R.~D.~Ball {\it et al.}  [The NNPDF Collaboration],
  Nucl.\ Phys.\ B {\bf 855} (2012) 153
  [arXiv:1107.2652 [hep-ph]].


\bibitem{HERAPDF15}
  H1 and ZEUS Collaborations,
  {H1prelim-10-142, ZEUS-prel-10-018}.


\bibitem{D'Agostini:1993uj}
  G.~D'Agostini,
  Nucl.\ Instrum.\ Meth.\ A {\bf 346} (1994) 306.

\bibitem{Blobel:2003wa}
  V.~Blobel,
  {\it In the Proceedings of PHYSTAT2003}. 

\bibitem{Abazov:2009nc}
  V.~M.~Abazov {\it et al.}  [D0 Collaboration],
  Phys.\ Rev.\ D {\bf 80} (2009) 111107
  [arXiv:0911.2710 [hep-ex]].


\bibitem{Bethke:2011tr}
  S.~Bethke {\it et al.}, Proceedings of the ``Workshop on Precision Measurements of $\alpha_s$'',
  arXiv:1110.0016 [hep-ph].

\end{thebibliography}








\end{document}